\documentclass[pre,aps,floatfix,twocolumn,preprintnumbers,superscriptaddress,notitlepage,longbibliography]{revtex4-1}
\usepackage{graphicx}
\usepackage{dcolumn}
\usepackage{bm}
\usepackage{amssymb}
\usepackage{pifont}
\usepackage{amsmath}
\usepackage{times}
\usepackage{natbib}

\usepackage{epstopdf}
\usepackage{latexsym}
\usepackage{keyval}
\usepackage{ifthen}
\usepackage{moreverb}
\usepackage{color}

\def \ii {i}

\def \fphi {\mathcal{G}}
\def \fr {\mathcal{F}}

\newcommand{\eqqref}[1] {~(\ref{#1})}

\DeclareMathOperator{\ee}{\rm e}

\DeclareMathOperator{\Real}{Re}
\DeclareMathOperator{\Imag}{Im}

\begin{document}
\title{From phase to amplitude oscillators}
\author{Pau Clusella}
\affiliation{Institute for Complex Systems and Mathematical Biology, SUPA,
	University of Aberdeen, Aberdeen AB24 3UE, United Kingdom}
\affiliation{Dipartimento di Fisica, Universit\`a di Firenze, I-50019 Sesto Fiorentino, Italy}

\author{Antonio Politi}
\affiliation{Institute for Complex Systems and Mathematical Biology, SUPA,
	University of Aberdeen, Aberdeen AB24 3UE, United Kingdom}
\date{\today}

 \begin{abstract}
We analyze an intermediate collective regime where amplitude oscillators 
distribute themselves along a closed, smooth, time-dependent curve $\mathcal{C}$,
thereby maintaining the typical ordering of (identical) phase oscillators.
This is achieved by developing a general formalism based on 
two partial differential equations, which describe 
the evolution of the probability density along $\mathcal{C}$ and of the shape of $\mathcal{C}$ itself.
The formalism is specifically developed for Stuart-Landau oscillators, 
but it is general enough to be applied to other classes of amplitude oscillators.
The main achievements consist in: 
(i) identification and characterization of a new transition to self-consistent
partial synchrony (SCPS), which confirms the crucial role played by 
higher Fourier hamonics in the coupling function;
(ii) an analytical treatment of SCPS, including a detailed stability analysis; 
(iii) the discovery of a new form of collective chaos, which can be seen as a 
generalization of SCPS and characterized by a multifractal probability density.
Finally, we are able to describe given dynamical regimes both at the macroscopic as well as 
the microscopic level, thereby shedding further light on the relationship
between the two different levels of description.
\end{abstract}

\maketitle 

\section{Introduction}
One of the peculiarities of high-dimensional complex systems is the spontaneous emergence of
nontrivial dynamical regimes over multiple scales.
The simultaneous presence of a macroscopic and a microscopic dynamics in mean-field models is 
perhaps the simplest instance of this phenomenon and, yet, it is not fully understood.
It is, for instance, not clear how and to what extent the microscopic and macroscopic world are
related to one another; even the minimal properties required for the emergence of a non trivial
collective dynamics are unknown\footnote{The term ``collective'' is often used to refer to the behavior 
of an ensemble of units induced by mutual coupling.
Here we use it to refer, as in statistical-mechanics, to macroscopic features exhibited by observables averaged over a formally infinite number of contributions.
}.
This difficulty is not a surprise, as the problem is akin to the emergence of different 
thermodynamic phases in equilibrium statistical mechanics, with the crucial difference that 
there are no Boltzmann-Gibbs weights that can be invoked and, moreover, the thermodynamic phases 
themselves are not steady. 

Generally speaking, collective regimes in mean-field models can be classified into two large
families: (i) symmetry-broken phases, where the ensemble elements split in two or more groups, 
each characterized by its own dynamics; (ii) symmetric phases, where all oscillators 
behave in the same way.
Clustered~\cite{Golomb-etal-92} and chimera~\cite{Kuramoto-Battogtokh-02,Abrams-Strogatz-04} states are the 
two most prominent examples of the former type, while fully synchronous and splay states are the simplest
examples of the latter one.
In this paper we focus on the latter class and, more precisely, on the emergence of collective
chaos, our goal being to shed light on the way macroscopically active degrees of freedom
may spontaneously emerge in an ensemble of identical units.

It is natural to classify the different setups
according to the dynamical complexity of the single oscillators, since the overall behavior should be
ultimately traced back to the rules determining the evolution of the single elements.
Intrinsically chaotic elements such as logistic maps sit at the top of the hierarchy.
It is known since many years that collective chaos may sponentaneously emerge as a result of subtle
correlations among the single dynamical units~\cite{Kaneko-90}.
Collective chaos can emerge also in periodic oscillators such as the Stuart-Landau model~\cite{nakagawa1993,hakim,chabanol}, since the
macroscopic mean field can trigger and maintain a microscopic chaos, thereby giving rise to a regime
similar to the previous one.

By further descending the ladder of the single-unit complexity, no evidence of collective chaos has been found
in mean-field models of identical phase-oscillators, but no rigorous argument excluding that this can
happen is available. In fact, in all mean-field systems of identical oscillators, the probability distribution
satisfies a self-consistent functional equation which, being nonlinear and infinite-dimensional can in principle 
give rise to a chaotic dynamics.
Nevertheless, collective chaos in phase oscillators has been found only when either 
delayed interactions~\cite{Pazo-Montbrio2016}, multiple populations~\cite{Olmi-etal-10},
or heterogeneity~\cite{Luccioli-Politi-10} is included in the model.
Otherwise, at most macroscopic periodic oscillations arise (the so-called self-consistent partial synchrony 
SCPS)~\cite{vanVreeswijk1996,Mohanty-Politi-06,Rosenblum-Pikovsky-2007,Pikovsky-Rosenblum-2009,
Politi-Rosenblum-2015, Rosenblum-Pikovsky-15}, the 
simplest setup for their observation being the biharmonic Kuramoto-Daido model~\cite{Clusella-Politi-Rosenblum2016}.
The so-called balanced regime observed in the context of neural networks is, instead, an entirely 
different story, since the collective dynamics is basically the result of the amplification of 
microscopic fluctuations~\cite{Ullner-etal-18}.

Altogether, new theoretical approaches are required to improve our general understanding of the collective
behavior of ensembles of dynamical units.
In phase oscillators characterized by strictly sinusoidal coupling-functions (Kuramoto type) the
Watanabe-Strogatz theorem~\cite{Watanabe-Strogatz-93} and the Ott-Antonsen Ansatz~\cite{Ott-Antonsen-2008} imply that no more than three collective degrees of freedom can emerge. What can we say in more general contexts?
In this paper, we discuss an intermediate regime, where amplitude oscillators (characterized by at
least two variables) retain a key property of phase oscillators, \emph{i.e.} they can be parametrized by a single 
phase-like variable.  In practice, the amplitude dynamics is sufficiently stable to force the oscillators towards
a smooth closed curve $\mathcal{C}$ such as in standard phase oscillators, but not stable enough to prevent fluctuations 
of the curve itself, which therefore acts as an additional source of complexity.
We refer to this regime, which has been so far basically overlooked, as to Quasi Phase Oscillators (QPO). 
We are only aware of a preliminary study carried out by Kuramoto \& Nakagawa, who performed microscopic
simulations of an ensemble of Stuart-Landau oscillators~\cite{nakagawa1995}.
Here we develop a general formalism, which allows describing QPO at a macroscopic level.
As a result, we are able to identify and describe a series of dynamical regimes ranging from plain SCPS to 
a new type of collective chaos.
SCPS itself is already an intriguing regime. In fact, under the action of a self-consistent mean field,
the single oscillators exhibit generically a quasiperiodic behavior without phase locking (no Arnold tongues).
Said differently, the transversal Lyapunov exponent of each single oscillator (conditioned to a given ``external''
forcing) is consistently equal to zero, when a control parameter is varied.
The hallmark of QPO is that the transverse Lyapunov exponent is not positive, even when the ``external'' mean-field
forcing is chaotic: actually, it is even slightly negative, implying that the phase probability distribution 
is not smooth but multifractal.
This is at variance with the ``standard'' collective chaos observed in a different parameter region
of the Stuart-Landau model~\cite{nakagawa1993,Takeuchi-Chate2013,hakim,chabanol},
characterized by a positive transversal Lyapunov exponent.

In Section~\ref{sec:model} we introduce the model and develop the formalism, which consists in the derivation of
two partial differential equations (PDEs), one describing the probability density of the oscillators
along the curve $\mathcal{C}$ and the other describing the shape of the curve itself.
These equations are thereby simulated to identify the different regimes: the agreement with the 
integration of the microscopic equations validates the whole approach. 
Next, an exact stability analysis of the splay state is performed in Section~\ref{sec:splay}, which allows
determining the critical value of the coupling strength where SCPS is born.
Section~\ref{sec:SCPS} is devoted to the analysis of SCPS, leading to an exact solution for the distribution
of phases and for the shape of $\mathcal{C}$. Therein we analyse also the transition to SCPS, uncovering very
subtle properties: superficially, the transition corresponds to the critical point of the Kuramoto-Sakaguchi 
model~\cite{Sakaguchi-Kuramoto-86}, 
which separates the stability of the splay state from that of the synchronous solution. 
However, this is not entirely correct, as it is known that the Kuramoto-Sakaguchi model cannot support SCPS.
In fact, it turns out that the mutual coupling and, in particular, the fluctuations of the curve $\mathcal{C}$,
give birth to higher Fourier harmonics, which are essential for the stabilization of SCPS. 
In the following section~\ref{sec:chaos}, we analyse the increasingly complex regimes generated when the coupling
is further increased. In particular, we see that the transition to chaos is rather anomalous: a sort of
intermediate scenario between the standard low-dimensional route to chaos where a single exponent becomes 
positive, and high-dimensional transitions when the number of unstable directions is proportional to the
system size. In the present context, we have evidence of the simultaneous appearance of 
three positive exponents, irrespective of the number of oscillators.
An additional unusual feature of the chaotic dynamics is the presence of (infinitely many) nearly zero exponents, 
which, according to the Kaplan-York formula, make the system high dimensional in spite of the finite (and small)
number of unstable directions.
In section~\ref{sec:chaos} we discuss also macroscopic and microscopic Lyapunov exponents, showing
that they are consistent with one another. This equivalence is far from trivial, since in all previous instances
of collective chaos the correspondence between the two spectra was at most restricted to very
few exponents, the reason for the difference being that while the microscopic exponents refer to
perturbations of single trajectories, the macroscopic ones refer to perturbations of the distributions
of phases.
Finally, section~\ref{sec:conclusions} contains a summary of the main results and of the open problems.

\section{The model}\label{sec:model}
Following Ref.~\cite{nakagawa1993}, we consider an ensemble of $N$ coupled Stuart-Landau oscillators
\begin{equation}\label{eq:coupled-oscillators}
	\dot z_j=K(1+ic_1)(\overline z-z_j)+ z_j-(1+ic_2)|z_j|^2 z_j
\end{equation}
where $z\in\mathbb{C}$ and
\begin{equation*}
	\bar z=\frac{1}{N}\sum_{m=1}^N z_m
\end{equation*}
is the mean field.
The formulation of Refs.~\cite{chabanol, hakim} 
is recovered by setting $t\to Kt$, $\alpha=c_2$, $\eta=c_1$ and $\mu=K^{-1}$.\\

A homogeneous system of globally coupled oscillators such as~(\ref{eq:coupled-oscillators}) can 
display two types of stationary solutions: a state of \emph{full synchrony} and \emph{incoherent states}.
In the state of full synchrony all the oscillators evolve periodically as 
$z_j=\overline z=\mathrm{e}^{i c_2 t}$.  
Incoherent states correspond to a distribution of oscillators 
such that the mean field $\overline z$ vanishes. There are infinitely many ways of realizing
an incoherent state, the simplest being a completely homogeneous distribution of the phases
along a circle (of radius $\sqrt{1-K}$) also known as \emph{splay state}.

For small coupling strength $K\ll 1$, the model can be mapped onto a Kuramoto-Sakaguchi model~\cite{Rosenblum-Pikovsky-15},
which is known to give rise only to either fully synchronous or splay states.
For larger coupling strength, the model displays a rich variety of dynamics, 
including clustered states, chimera states, and different forms of collective 
chaos~\cite{nakagawa1993,nakagawa1995,chabanol,hakim,sethia}.
In particular, Nakagawa \& Kuramoto \cite{nakagawa1995}  reported numerical evidence of a dynamical regime where
the oscillators seem to be distributed along a smooth time-dependent closed curve.
In this paper, we revisit such a behavior, developing a formalism, which can, in principle,
be applied to generic ensembles of mean-field driven amplitude oscillators.

\begin{figure}
	\includegraphics[width=0.4\textwidth]{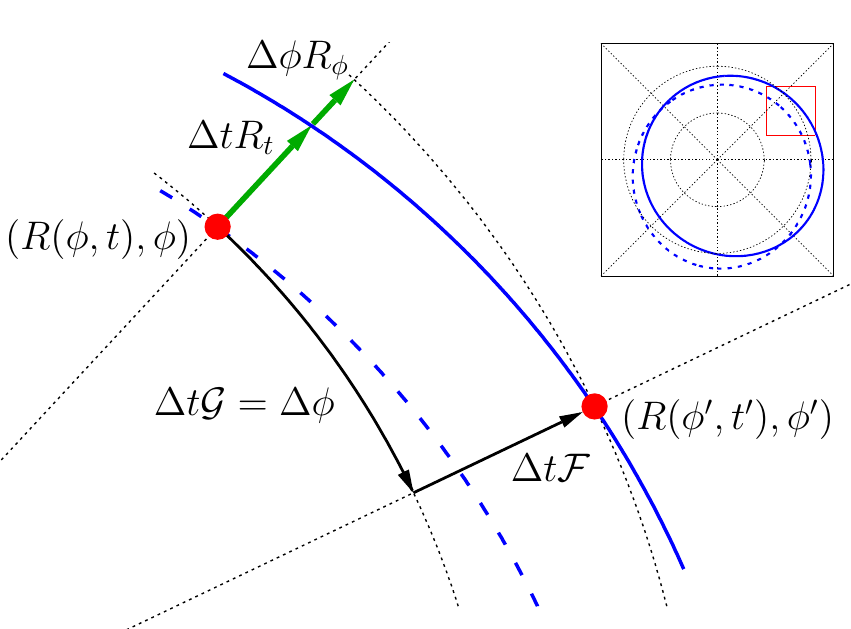}
	\caption{Schematic representation of the phase parametrization of $\mathcal{C}$ and its time evolution.
	Dashed blue curve depicts the shape of of $\mathcal{C}$ at time $t$, whereas the continuous blue
	line indicates the position of $\mathcal{C}$ at time $t'=t+\Delta t$. The inset shows a zoom out version,
	with the red rectangle indicating the zone depicted in the main figure.
	}\label{fig:description}
\end{figure}
A general macroscopic formulation of system~\eqref{eq:coupled-oscillators}
requires the knowledge of the probability density on the complex plane $Q=Q(z,t)$. 
Nevertheless, if the amplitude dynamics is sufficiently stable, the oscillator
variables naturally converge and are eventually confined to a closed curve $\mathcal{C}$. 
Whenever this is the case, $\mathcal{C}$ can be uniquely parametrized in polar coordinates by expressing the
radius $R(\phi,t)$ as a function of the phase $\phi$.
Accordingly we can assume that the probability density is confined to $\mathcal{C}$ so that
\begin{equation*}
Q(z,t)=Q(z,t)\delta(|z|-R(\phi,t))=:P(\phi,t).
\end{equation*}
The meaningfulness of this approach is verified a posteriori by the consistency of the theoretical
predictions and by the agreement with direct numerical simulations. 

Our first goal is to derive the evolution equations for $R$ and $P$.
In order to do so, it is convenient to express Eq.~(\ref{eq:coupled-oscillators})
in polar coordinates, $z=r\ee^{\ii\phi}$, 
\begin{align*}
\dot r_j&=  \fr[r_j,\phi_j,\overline z]\nonumber \\ 
\dot \phi_j&= \fphi[r_j,\phi_j,\overline z] \;,
\end{align*}
where
\begin{align*}
\fr[r,\phi,\overline z] &= (1-K- r^2)r+K\Real[(1+i c_1)\overline z \ee^{-i\phi}]\nonumber \\ 
\fphi[r,\phi,\overline z] &= -c_1 K-c_2 r^2+\frac{K}{r}\Imag[(1+i c_1)\overline z\ee^{-i\phi}] \;.
\end{align*}
Let us now consider a time interval $\Delta t\ll 1$ and $t'=t+\Delta t$.
At time $t$ the oscillator with index $j$ has polar coordinates $[r_j(t),\phi_j(t)]=[R(\phi,t),\phi]$
while at time $t'$ has $[r_j(t'),\phi_j(t')]=[R(\phi',t'),\phi']$, as indicated in figure~\ref{fig:description}.
If $R$ is smooth we have that
\begin{equation*}
\begin{cases}
R(\phi', t')=R(\phi,t)+\Delta t\;  \fr[R(\phi,t),\phi,\overline z]\\
\phi'=\phi+\Delta t \; \fphi[R(\phi, t), \phi, \overline z].
\end{cases}
\end{equation*}
Then one can write
\begin{equation*}
\frac{R(\phi,t')-R(\phi,t)}{\Delta t} =
\frac{R(\phi',t')-R(\phi,t)}{\Delta t}-\frac{\phi'-\phi}{\Delta t}R_\phi
\end{equation*}
so that
\begin{equation}\label{eq:attractor2}
\frac{\partial R}{\partial t}(\phi,t)=
\fr[R,\phi,\overline z]-\fphi[R,\phi, \overline z]R_\phi \; .
\end{equation}
On the other hand, the evolution of $P$ is determined by the angular flux at $r=R(\phi,t)$,
\begin{align}\label{eq:probability}
\frac{\partial P}{\partial t}(\phi,t)&=
-\frac{\partial }{\partial \phi}\Bigl\{
P(\phi,t) \fphi[R,\phi,\overline z]
\Bigr\}.
\end{align}
while the mean field is finally defined as
\begin{equation}\label{eq:coupling1}
\overline z(t)=\int_0^{2\pi} P(\phi,t)R(\phi,t)e^{i\phi}d\phi.
\end{equation}\\

From the expression for $\fphi[R,\phi,\overline z]$, one can recognize the 
Kuramoto-Sakaguchi structure~\cite{Sakaguchi-Kuramoto-86} of the velocity field, with however, the important
difference of the additional factor $R(\phi,t)$ in the definition of the
order parameter. We will see that the time dependence of $R(\phi,t)$ (it obeys a distinct
differential equation) enriches the complexity of the collective dynamics.

Overall, equations (\ref{eq:attractor2},\ref{eq:probability},\ref{eq:coupling1}) represent a system 
of two nonlinear Partial Differential Equations (PDEs) describing the macroscopic behavior of the oscillators 
whenever they are spread along a closed phase-parameterized smooth curve.

Such a system can be solved numerically by means of a split-step Fourier or pseudospectral method~\cite{Canuto2006}.
The algorithm consists in expanding $R$ and $P$ spatially in Fourier space,
\begin{align*}
P(\phi,t)&=\frac{1}{2\pi}\sum_{k=-\infty}^\infty \tilde P(k,t)e^{-ik\phi}\quad\text{and} \nonumber \\
R(\phi,t)&=\frac{1}{2\pi}\sum_{k=-\infty}^\infty \tilde R(k,t)e^{-ik\phi}
\end{align*}
where
\begin{align*}
\quad \tilde P(k,t)&=\int_0^{2\pi}d\phi\; P(\phi,t)e^{ik\phi}\quad\text{and} \nonumber\\
\quad \tilde R(k,t)&=\int_0^{2\pi}d\phi\; R(\phi,t)e^{ik\phi}.
\end{align*}
By truncating the Fourier series for a large enough wavelength $M$,
one can then integrate in time the different Fourier modes $\tilde P(k,t)$ and $\tilde R(k,t)$ using
a standard method for ordinary differential equations.
Since the computation of a nonlinear term of order $q$ in Fourier space requires $\mathcal{O}(M^q)$
operations, it is more convenient to compute the velocity fields in real space instead.
In fact, by invoking fast Fourier transform (FFT) algorithms, the computational cost of the nonlinear 
fluxes reduces to $\mathcal{O}(4n\log(2n))$, where $n$ is the number of points of the real support.
We use a fourth-order Runge-Kutta integration method with time step $\delta t$, set to $10^{-3}$ in 
most of the simulations.
In order to avoid aliasing errors in the successive calls of the FFT 
 we use the 3/2-truncation rule using a real spatial grid of $n=2048$ points and a
truncation of $M=680$ in the Fourier expansion.
Nevertheless, in the presence of chaotic dynamics (see below) the numerics become more challenging
and it is necessary to increase both spatial and temporal resolutions.
Therefore, for $K>0.416$ we use $M=1024$ and $\delta t =0.5\cdot 10^{-3}$, and keep $n=2048$.
In this case it is necessary to introduce a smoothing technique to prevent the growth of the aliasing errors.
Thus we add a small diffusive term in equation~(\ref{eq:probability}),
\begin{align*}
\frac{\partial P}{\partial t}(\phi,t)&=
-\frac{\partial }{\partial \phi}\Bigl\{
P(\phi,t) \fphi[R,\phi,\overline z]
\Bigr\}+D\frac{\partial^2}{\partial \phi^2}P(\phi,t)\;.
\end{align*}
A diffusion level of $D=10^{-8}$ is enough to stabilize the algorithm while preserving the dynamical properties of the system.
\\

\begin{figure}
	\includegraphics[width=0.5\textwidth]{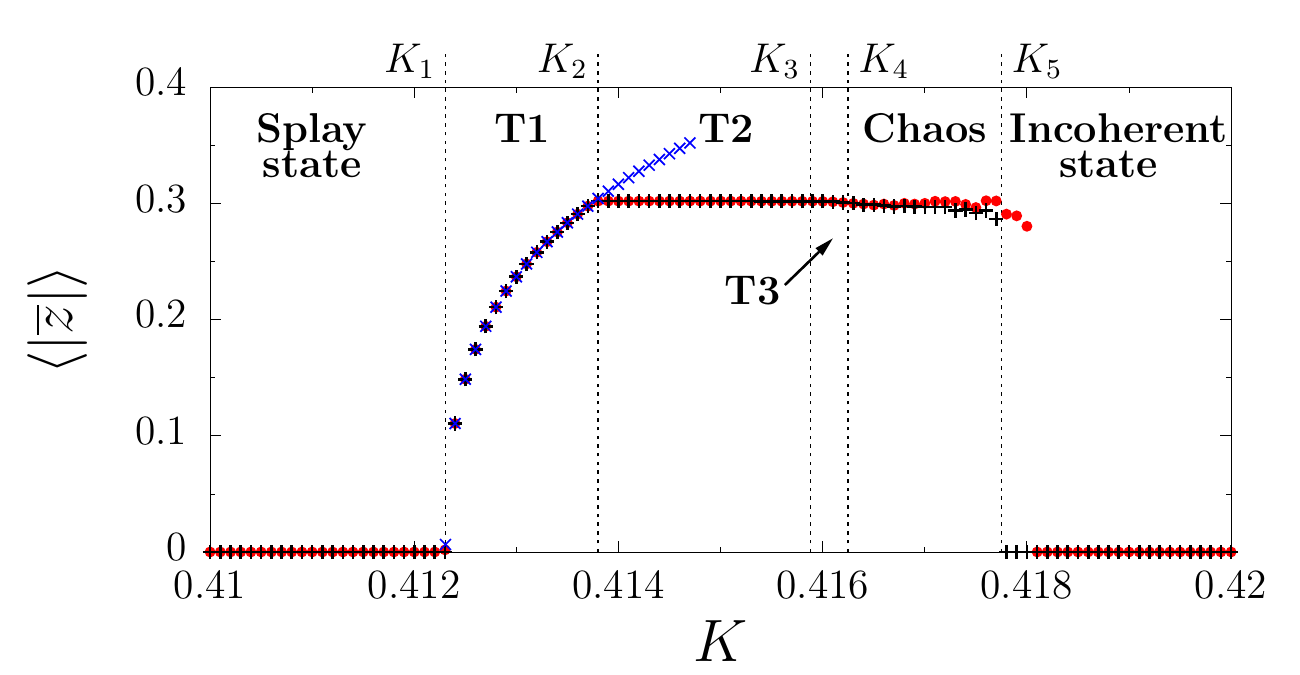}
	\caption{Time-averaged mean-field $\langle |\overline{z}|\rangle$ for different values of the coupling strength $K$ and fixed $c_1=-2$ and $c_2=3$.
		Black plusses show the results from microscopic simulations of the system as given by \eqref{eq:coupled-oscillators} with $N=4096$ oscillators computed over $10^6$ time units after discarding another $10^6$ of transient.
		Red circles correspond to the numerical integration of Eqs.~(\ref{eq:attractor2}) and~(\ref{eq:probability}) with a pseudospectral method for $10^5$ time units, from which $5\cdot 10^4$ are transient.
		Both microscopic and macroscopic simulations use initial conditions close to splay state.
		Blue crosses correspond to the semi-analytical fixed point given by Eqs.~(\ref{eq:scpsR}) and~(\ref{eq:scpsP}).
		Black vertical dashed lines indicate the bifurcation points between the different dynamical regimes.		
	}\label{fig:phasediagram}
\end{figure}

\begin{figure}
	\includegraphics[width=0.5\textwidth]{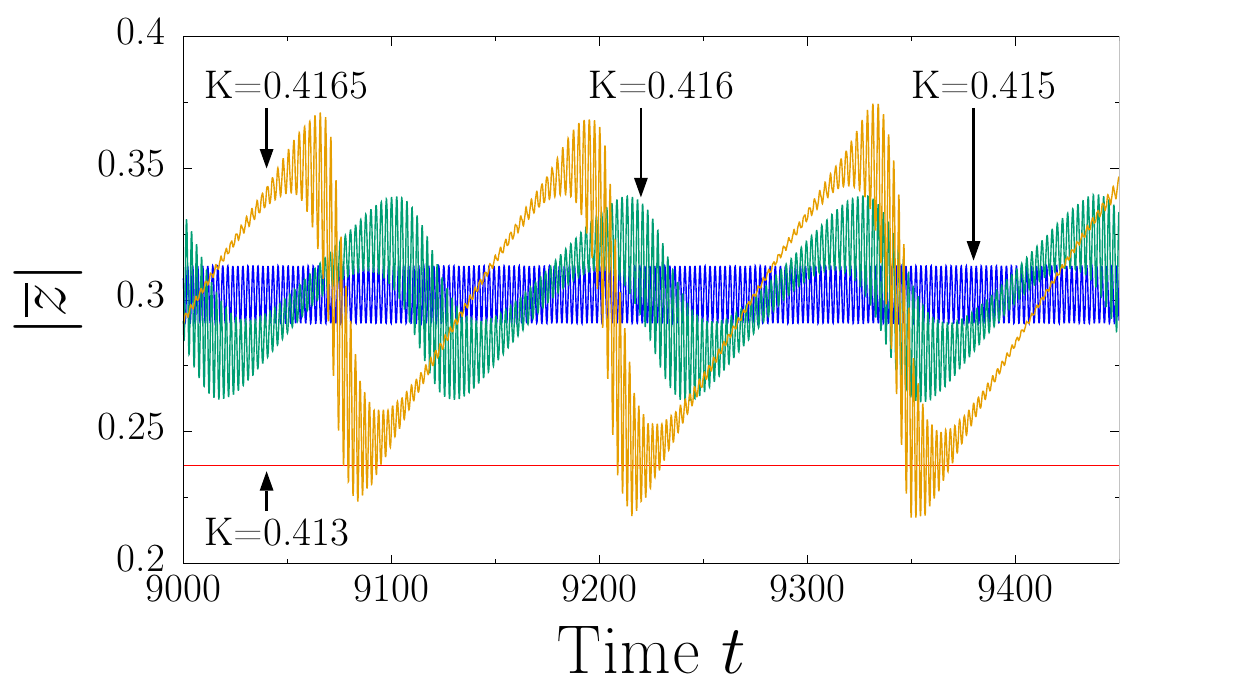}
\caption{Time series of $|\overline z|$ from the numerical integration of the macroscopic equations of the system
	for $K=0.413$ (red), $0.415$ (blue), $0.416$ (green), and $0.4165$ (amber).}
\label{fig:timeseries}
\end{figure}

\begin{figure*}
	\includegraphics[width=1\textwidth]{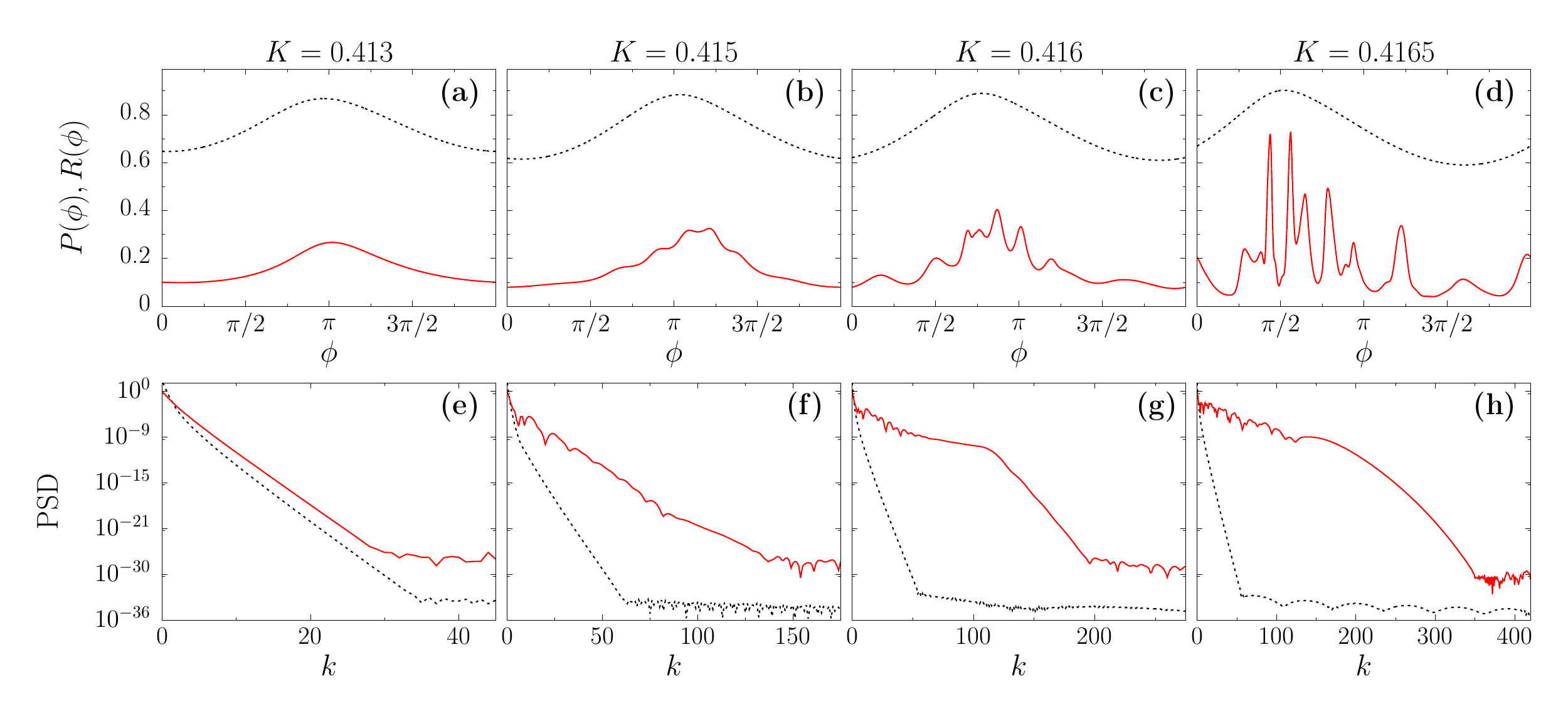}
	\caption{Snapshots of the shape of $R$ (dashed black curve) and $P$ (red continuous curve) in real space (top figures), and the
		corresponding Power Spectral Density in logarithmic scale (bottom figures).
		(a) and (e) correspond to $K=0.413$, (b) and (f) to $K=0.415$, (c) and (g) to $K=0.416$, and (d) and (h) to $K=0.4165$.
		Results obtained upon integrating the macroscopic equations using a pseudospectral method.
	}
	\label{fig:snapshots}
\end{figure*}

In order to double check our numerics we have fixed two parameters as in Ref.~\cite{nakagawa1995}
($c_1=-2$ and $c_2=3$) and treated the coupling strength $K$ as the leading control parameter.
The results of our numerical simulations are reported in 
figures~\ref{fig:phasediagram}, \ref{fig:timeseries} and~\ref{fig:snapshots}.
In figure~\ref{fig:phasediagram} we show the time-average of the mean-field $|\overline z|$ computed both 
using the microscopic formulation of the system~(\ref{eq:coupled-oscillators}) with $N=4096$ (black plusses) 
and using the macroscopic equations solved with the pseudospectral method (red circles).  
The nice agreement confirms the correctness of the macroscopic equations.

The behavior of the order parameter unveils a series of bifurcations across various dynamical regimes, 
marked with vertical dashed black lines. 
In figure~\ref{fig:timeseries} we show the time evolution of the order parameter modulus $|\overline z|$ for different values of $K$, 
each belonging to a different dynamical regime. 
In the top panels of figure~\ref{fig:snapshots} we plot snapshots of $R$ (black dashed lines) and the 
corresponding probability densities $P$ (red lines). While the shape of the curve $\mathcal{C}$ 
remains very smooth when the coupling is increased, the probability profiles become increasingly rugged,
revealing an increasing contribution of higher Fourier modes. This is better appreciated by looking
at the bottom panels of Fig.~\ref{fig:snapshots}, where the corresponding Power Spectral Density 
(PSD) is reported in a logarithmic scale (same color code for the different curves).
\\

For $K<K_1\approx 0.4123$, the oscillators are uniformly distributed on a perfect circle and the mean-field 
$\overline z$ vanishes, \emph{i.e.}, the system is in a \emph{splay state}. 
For $K=K_1$ there is a first transition to a regime where the mean-field dynamics is a pure periodic rotation 
on a one-dimensional torus (T1), accompanied by a quasiperiodic dynamics of the single oscillators.
This regime is therefore an example of \emph{self-consistent partial synchrony} (SCPS) first uncovered in a model
of leaky integrate-and-fire neurons~\cite{vanVreeswijk1996} and fully described in biharmonic Kuramoto-Daido
oscillators~\cite{Clusella-Politi-Rosenblum2016}. In this regime, the mean-field modulus is constant 
(see red line in Fig.~\ref{fig:timeseries}). In fact, both $P$ and $R$ are spatially nonuniform
steady functions if observed in a suitably rotating frame 
(see Fig.~\ref{fig:snapshots}(a), where we can also notice that the shape is dominated by a few long wavelength
modes - panel (e)). 
At $K=K_2\simeq 0.4138$ a macroscopic Hopf bifurcation occurs, which introduces a second frequency. As a result, the
collective dynamics is quasiperiodic (T2, but periodic, if observed in a suitably rotating frame)
(see the blue line in Fig.~\ref{fig:timeseries}). Now the shape of $P$ and $R$ fluctuate in time and more Fourier modes
contribute 
(see figures~\ref{fig:snapshots}(b) and (f)).

At $K=K_3 \simeq 0.41588$, yet another frequency adds to the macroscopic dynamics, although
indistinguishable from the average value of the order parameter.
Thus, as better argued in the following, the dynamics of the order parameter becomes three-dimensional (T3) 
(see the green line in Fig.~\ref{fig:timeseries}).
The shape of $P$ becomes more uneven, with several bumps that evolve on time, and the spatial 
spectra involve higher Fourier modes (see Fig.~\ref{fig:snapshots}(c) and (g)).

The T3 regime is stable for a small range of the parameter value; 
beyond $K=K_4 \simeq 0.41605$ the system becomes chaotic.
Although $R$ keeps being quite smooth, the shape of $P$ occasionally develops several peaks that 
become sharper upon increasing the coupling strength (see Fig.~\ref{fig:snapshots}(d) and (h)).
Such peaks generate spurious ringing artifacts that eventually lead to numerical instabilities.
Nevertheless, as explained before, increasing the numerical resolution and introducing a negligible 
diffusion, one can accurately integrate the macroscopic equations for long times.

For yet larger $K$ values, the order parameter vanishes after a long chaotic transient.
However, the asymptotic regime is not a splay state: only the first Fourier mode vanishes, while
the others have non zero amplitude. In other words the distribution of angles is non uniform.
The disagreement between microscopic and macroscopic simulations on the bifurcation point $K_5$ 
in figure~\ref{fig:phasediagram} is due to the fact that macroscopic simulations cannot be run
long enough for the chaotic transient to finish.
Finally, if the coupling is increased even further, one ends up in the highly irregular chaotic
regime studied in~\cite{nakagawa1993,Takeuchi-Chate2013}, where
the macroscopic equations~(\ref{eq:attractor2}) and~(\ref{eq:probability}) are no longer valid,
since the oscillators are not distributed along a smooth closed curve.

\section{Stability of Splay State}\label{sec:splay}

The macroscopic equations can be used to  perform the stability analysis of the splay state by 
introducing the fields $u(\phi,t)$ and $v(\phi,t)$ which denote infinitesimal perturbations of the
probability $P_0=1/(2\pi)$ and of the radius $R_0=\sqrt{1-K}$, respectively.
Upon linearizing Eqs.\eqqref{eq:attractor2} and \eqqref{eq:probability}, it is found that
\begin{equation*}
v_t=-2(1-K)v+A v_\phi + K \Real[ (1+ic_1)\overline w \mathrm{e}^{-i\phi} ] \;, 
\end{equation*}
and
\begin{align*}
u_t=&A u_\phi+\frac{c_2\sqrt{1-K}}{\pi}v_\phi
+\frac{K\Real[ (1+ic_1)\overline w \mathrm{e}^{-i\phi} ]}{2\pi \sqrt{1-K} } \;,
\end{align*}
where
\begin{align*}
A =& c_1 K+c_2(1-K)
\end{align*}
and
\begin{align*}
\overline w:=&
\sqrt{1-K}\int_0^{2\pi}\!\!\!d\phi\; u(\phi,t)\ee^{i\phi}+
\frac{1}{2\pi}\int_0^{2\pi}\!\!\!d\phi\; v(\phi,t)\ee^{i\phi}
\end{align*}
is the (linear) variation of the mean-field. It is composed of two terms: the former one is the standard 
Kuramoto-type order parameter, while the latter accounts for fluctuations  of the curve $\mathcal{C}$.

The linearized equations can be easily solved in Fourier space, \textit{i.e.}, by introducing 
\begin{align*}
\tilde v(k,t)&=\int_{0}^{2\pi}d\phi  v(\phi,t)\ee^{i k\phi}\qquad\text{and}\nonumber\\
\tilde u(k,t)&=\int_0^{2\pi}d\phi  u(\phi,t)\ee^{i k\phi}\;,
\end{align*}
(and the corresponding inverse transforms), since they become block diagonal.

For $k\neq\pm1$ (including $k=0$),
\begin{align*}
[\tilde v(k)]_t&=\left[ -2(1-K)-\ii k A \right]\tilde v(k)\quad\text{and} \nonumber\\
[\tilde u(k)]_t&=-\ii k \frac{c_2\sqrt{1-K}}{\pi}\tilde  v(k)-\ii k A\tilde u(k) \;
\end{align*}
so that, the evolution of $\tilde v(k)$ is closed onto itself and acts as an external forcing for the 
dynamics of the probability density, yielding the eigenvalues
\begin{align*}
\lambda^{(v)}_k&=-2(1-K)-\ii k (Kc_1+c_2(1-K))\quad\text{and}\\
\lambda^{(u)}_k&=-\ii k\left(Kc_1+c_2(1-K)\right) \; .
\end{align*}
Altogether, the eigenvalues are arranged into two branches: the former corresponds to stable directions 
associated to the $\mathcal{C}$-dynamics; the latter corresponds to marginally stable directions associated to 
the density dynamics (this includes the zero mode, whose marginal stability
is nothing but a manifestation of probability conservation).
Overall the stability of the angle distribution is reminiscent of the marginal stability of the Kuramoto model.

The only exception is the first Fourier mode of $v(\phi,t)$, that is coupled back to the shape of the 
curve. For $k=1$ we obtain
\begin{align}\label{eq:decoupled}
[\tilde v(1)]_t&= M_{vv} \tilde v(1) + M_{vu} \tilde u(1) \\
[\tilde u(1)]_t&= M_{uv} \tilde v(1) + M_{uu} \tilde u(1) \nonumber 
\end{align}
where
\begin{align*}
M_{vv} =& -2 + \frac{5}{2}K-\ii\left(A-\frac{Kc_1}{2}\right) \nonumber \\
M_{vu} =& - \pi K\sqrt{1-K}(1+\ii c_1) \nonumber \\
M_{uv} =& \frac{K}{4\pi\sqrt{1-K}}-\ii\left( \frac{c_2\sqrt{1-K}}{\pi}-
\frac{Kc_1}{4\pi\sqrt{1-K}}\right) \nonumber \\
M_{uu} =& \frac{K}{2}-\ii\left(A-\frac{Kc_1}{2}\right)\; .
\end{align*}
Finally, the equations for $k=-1$ are obtained by simply taking
the complex conjugate of the above expressions.

Accordingly, instabilities can and actually do arise within the four-dimensional subspace spanned
by the real and imaginary parts of the first modes $\tilde v(1)$ and $\tilde u(1)$.
In practice one needs to determine the stability of the two dimensional complex system~(\ref{eq:decoupled}).
Using the Routh--Hurwitz criterion one finds two stability conditions
\begin{align*}
 &3K<2\;,\qquad\text{and}\\
&(2c_1^2+8c_1c_2-c_2^2+9)K^2 -\\&(c_1^2+12c_1c_2-c_2^2+12)K + 4 + 4c_1c_2 <0 
\end{align*}
In the parameter region considered in this paper, the second inequality reveals a loss of
stability for $K>K_1 = (65-\sqrt{1025})/80 = 0.412304\ldots$, 
which corresponds to a pair of complex conjugate eigenvalues crossing the imaginary axis, \emph{i.e.,}
a Hopf bifurcation. This bifurcation gives rise to SCPS, a periodic
collective dynamics, analyzed in the following section.

\section{Self-consistent partial synchrony}\label{sec:SCPS}
Above $K_1$, SCPS is stable within the region T1. Our general formalism, based on
Eqs.~(\ref{eq:attractor2},\ref{eq:probability},\ref{eq:coupling1}), allows performing an
analytical study of this regime. 

Self-consistent partial synchrony is characterized by a non-uniform probability density, 
rotating with a collective frequency $\omega$, which differs from the average
frequency of the single oscillators.
In models where the interactions depend only of phase differences like the present one, 
$\omega$ is constant as well as the amplitude $|\overline z|$ of the order parameter.
In other words, SCPS corresponds to a fixed point of Eqs.~(\ref{eq:attractor2},\ref{eq:probability}) 
in the moving frame $\theta=\phi-\omega t$.
Since we are free to choose the origin of the phases, we select a frame where $\overline z$ is real and positive,
 \emph{i.e.}, its phase vanishes
(so that from now on we can avoid the use of the absolute value).
The equations resulting from this change of variables are 
\begin{equation}\label{eq:rotating-frameR}
R_t= \fr[R,\theta,\overline z]+\left\{\omega-\fphi[R,\theta, \overline z]\right\}R_\theta
\end{equation}
and
\begin{equation}\label{eq:rotating-frameP}
P_t
=\frac{\partial }{\partial \theta}\Bigl(P(\theta,t) \bigl\{\omega-
 \fphi[R,\theta,\overline z]\bigr\}
\Bigr).
\end{equation}
accompanied by the self-consistent condition
\begin{equation}\label{eq:self-consistent}
\int_0^{2\pi}\!\!\!d\theta\; P(\theta,t)R(\theta,t)e^{i\theta}=:\overline z \; .
\end{equation}
By imposing the time derivative of $R$ equal to zero in equation~\eqqref{eq:rotating-frameR},
we find that the stationary shape of the limit cycle, $R_0(\theta)$, can be obtained by integrating the
following ODE,
\begin{equation}\label{eq:scpsR}
[R_0(\theta)]_\theta=-\frac{\fr[R_0,\theta,\overline z]}{\omega-\fphi[R_0,\theta, \overline z]}.
\end{equation}
Once a (numerical) expression for $R_0$ has been obtained, it can be replaced in the equation for $P$.
The stationary solution $P_0$ can be thereby obtained by setting the argument of the $\theta$ derivative
equal to a new unknown constant: minus the probability flux $\eta$. The solution reads
\begin{equation}\label{eq:scpsP}
P_0(\theta)=-\frac{\eta}{\omega-\fphi[R_0,\theta,\overline z]} \; ,
\end{equation}
where $\eta$ can be obtained from the normalization condition of the
probability density $\int_0^{2\pi}P_0=1$.
A complete expression for $R_0(\theta)$, and $P_0(\theta)$ is finally obtained by determining
the two remaining unknowns, $\overline z$ and $\omega$, from the complex self-consistent 
condition~\eqref{eq:self-consistent}. 
The resulting shapes are reported in Figs.~\ref{fig:scps}(a,b) for two different coupling strengths.
As expected, both $R_0$ and $P_0$ become increasingly nonuniform upon increasing the coupling strength.
Blue crosses in figure~\ref{fig:phasediagram} indicate the values of $|\overline z|$ obtained
from this approach.

In the rotating frame, where the curve $R(\theta)$ does not depend on time, one can view
the term in curly brackets in Eq.~(\ref{eq:rotating-frameP}) as the force field acting on an oscillator
of phase $\theta$, so that $\mathcal{G}(R(\theta),\theta)$ plays the role of the standard coupling term
in ensembles of phase oscillators.
The Kuramoto-Sakaguchi model is the simplest setup where the coupling is purely sinusoidal, \textit{i.e.},
$\mathcal{G} = G \sin(\psi +\alpha-\theta)$, where $G$ and $\psi$ are the amplitude and the phase of 
the order parameter, respectively. 
In such a case, it is well known that either the splay state or the fully 
synchronous solution are fully stable, depending whether $\alpha$ is larger or smaller than $\pi/2$.
Only for $\alpha=\pi/2$ intermediate macroscopically periodic solutions (SCPS-like) are possible,
and, moreover, the dynamics is highly degenerate, since infinitely many solutions exist and are thereby
marginally stable.
However, as soon as a small second harmonic is added, this high degree of degeneracy is lifted
and a finite parameter region appears, where robust SCPS can be observed (see~\cite{Clusella-Politi-Rosenblum2016}).
How do such findings compare with the scenario herein reported for QPOs? 

First of all, it should be reminded that the emergence of SCPS in Stuart-Landau oscillators
has been already reported in the past (see~\cite{Rosenblum-Pikovsky-2007,Rosenblum-Pikovsky-15}), 
a major difference being that
the theoretical and numerical studies refer to a parameter region where 
the single oscillators do not alter their phase character and the coupling manifests itself as a
nonlinear dependence on the order parameter and, last but not least, SCPS emerges as a loss
of stability of the fully synchronous state.

In the present context, the stability analysis of the splay state reveals that 
there is no need to include higher harmonics to correctly predict the onset of
SCPS, as if they were unnecessary. Since this result conflicts with our general understanding,
we have performed a perturbative expansion of the stationary solution in the vicinity of the critical point
(see appendix~\ref{sec:app2}).
The expansion implies that, at leading order, the coupling function $\mathcal{G}$ 
can be written as the sum of two terms, 
\begin{align*}
\mathcal{G} = 2c_2\sqrt{1-K} r(\theta)-K\bar z\sqrt{\frac{1+c_1^2}{1-K}}\sin(\nu-\theta) \; ,
\end{align*}
where $r$ is defined in Eq.~(\ref{eq:firstorder}) and $\nu =\arctan c_1$.
Since $r(\theta)$ is itself sinusoidal (see Eq.~(\ref{eq:perturbationr})),
$\mathcal{G}$ is purely sinusoidal as well.
The main difference with the Kuramoto-Sakaguchi model is that here there is an additional
indirect dependence of the order parameter through the modulation amplitude $r(\theta)$ of $\mathcal{C}$.
At criticality, the difference between the phase of the order parameter and that of
$r(\theta)$ is $\xi+\nu= -0.1736 $, while the difference with the phase of the probability density is
$\gamma+\nu=0.0523$.

At the same time, the perturbative analysis shows also that the amplitude $\bar z$ of the order 
parameter is undetermined at first order. 
This observation is consistent with the degeneracy exhibited by the Kuramoto-Sakaguchi model at criticality.
Moreover, it implies that a second harmonic needs to be included in the expansion
to obtain a full closure of the equations.
This is at variance with systems such as the biharmonic model studied in~\cite{Clusella-Politi-Rosenblum2016}, where
the shape of the probability density is characterized by the presence of a finite
second harmonics from the very beginning. This fact anyway confirms that
the presence of second (or higher) harmonics is crucial for the sustainment of
SCPS. In the present case such harmonics are spontaneously induced by the
modulation of $\mathcal{C}$.

\subsection{Stability analysis of SCPS}
Since $(R_0(\theta),P_0(\theta))$ is a fixed point of Eqs.~(\ref{eq:rotating-frameR}) and~(\ref{eq:rotating-frameP}),
one can easily study the stability of SCPS by determining the eigenvalues of the corresponding linear operator.
Let $v(\theta,t)$ and $u(\theta,t)$ denote an infinitesimal perturbation of $R_0(\theta)$ and $P_0(\theta)$, respectively.
By inserting $R(\theta,t)=R_0(\theta)+v(\theta,t)$ and $P(\theta,t)=P_0(\theta)+u(\theta,t)$ into 
Eqs.~(\ref{eq:rotating-frameR},\ref{eq:rotating-frameP})
and retaining only first order terms, we obtain the linear evolution equations,
\begin{align}\label{eq:scpsdotr}
 \left[v(\theta,t)\right]_t= v(\theta,t) & F^{(v)}(\theta)+\left[ v(\theta,t)\right]_\theta G^{(v)}(\theta)\\&+w(t)X^{(v)}(\theta)+\hat w(t)Y^{(v)}(\theta) \nonumber
\end{align}
\begin{align}\label{eq:scpsdotp}
 \left[ u(\theta,t)\right]_t=\frac{d}{d\theta}\Bigl[ v(\theta,t)&F^{(u)}(\theta)+u(\theta,t)G^{(u)}(\theta)\\&+w(t)X^{(u)}(\theta)+\hat w(t)Y^{(u)}(\theta)\Bigr] \nonumber
\end{align}
where  
\begin{align}\label{eq:selfconsistent}
w(t)&=\int_0^{2\pi}d\theta e^{i\theta}\left( vP_0+uR_0\right)\quad\text{and}\nonumber \\
\hat  w(t)&=\int_0^{2\pi}d\theta e^{-i\theta}\left( vP_0+uR_0\right) 
\end{align}
are the mean-field contributions in tangent space (see appendix~\ref{sec:app1} for the definition of the other
coefficients).
The linear equations are conveniently integrated in Fourier space even though (at variance with the
splay state) the change of variables does not diagonalize the problem. It is convenient to work in Fourier space,
since one can derive accurate solutions by truncating the infinite series (see the appendix), 
\emph{i.e.}, by neglecting all modes with $|k|>M$ for some suitably chosen value $M$.
The correct eigenvalues are thereby identified as those that are stable against an increase of
$M$. Additionally, the correctness of the selected values has been double checked by integrating the
corresponding eigenvectors (see Eqs.~(\ref{eq:scpsdotr},\ref{eq:scpsdotp})).
Not unexpectedly, the most relevant eigenvalues (and eigenvectors) do not require large $M$ values.

\begin{figure}
	\includegraphics[width=0.45\textwidth]{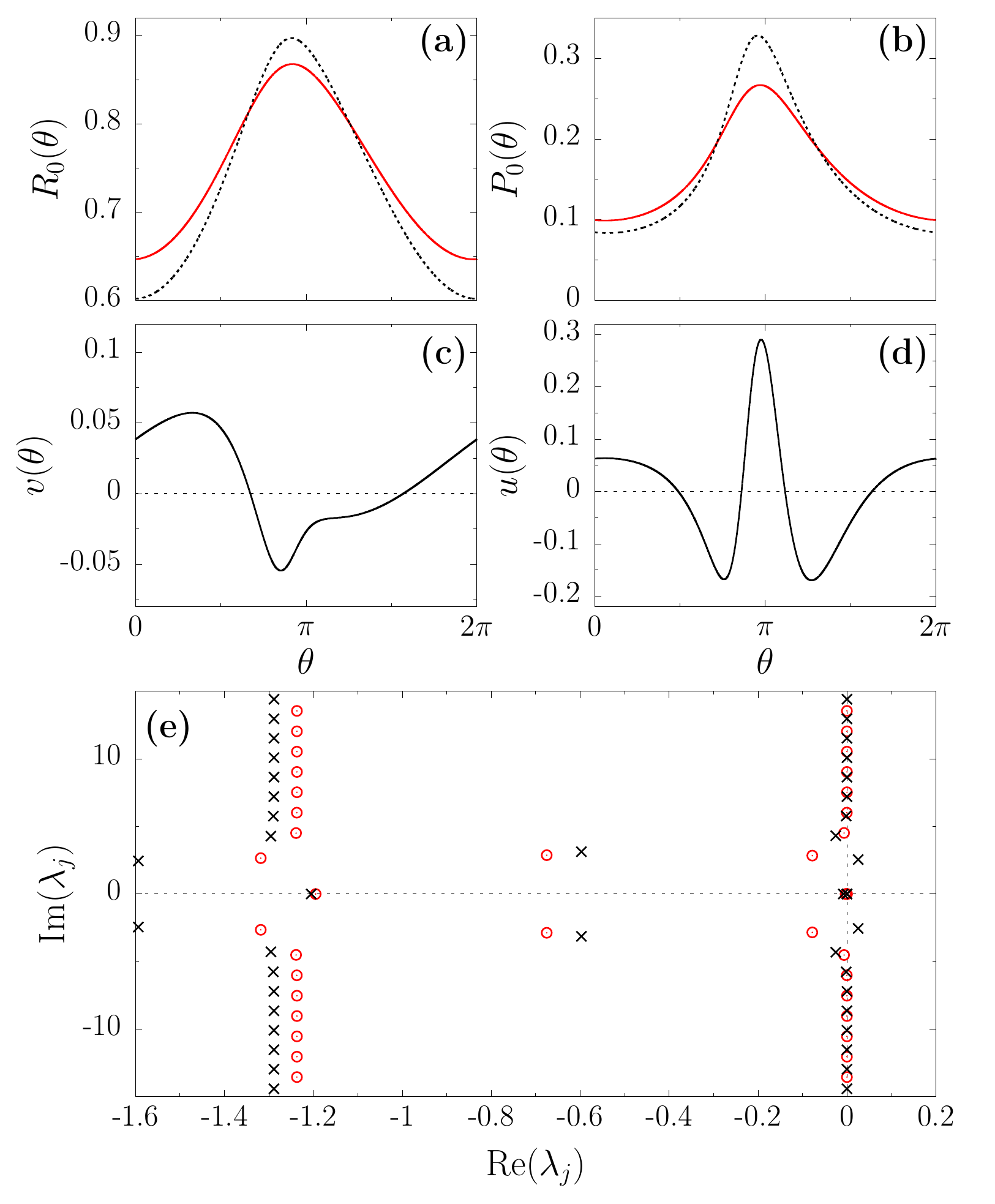}
	\caption{(a,b) Shapes of $R_0(\theta)$ and $P_0$ for $K=0.413$ (red) and $0.414$ (black dashed) obtained by solving Eqs.~(\ref{eq:scpsR},\ref{eq:scpsP}).
		(c,d) Shape of the unstable eigenfunction $v(\theta)$ and $u(\theta)$ for $K=0.414$. 
		(e) Eigenvalues resulting from diagonalizing the linear equations~(\ref{eq:scpsdotr}) and~(\ref{eq:scpsdotp}) with $M=400$ Fourier modes for $K=0.413$ (red circles) and $0.414$ (black crosses).}
	\label{fig:scps}
\end{figure}

In figure~\ref{fig:scps}(e) we show the resulting spectra for $K=0.413$ and $0.414$ ($M=400$ modes
have been used in the Fourier expansions). Analogously to the splay state, there are basically 
two sets of eigenvalues, one having strictly negative real parts, while the other corresponding 
to almost marginally stable directions (though not strictly vanishing as in the splay state).
For $K=0.413$ all the directions are stable, except for the one corresponding to the phase rotation,
while for $K=0.414$ a pair of complex conjugate eigenvalues is present with a positive real part.
The corresponding unstable eigenvectors are depicted in Fig.~\ref{fig:scps}(c,d).

Keeping track of such modes for different $K$ values, we can determine the 
second bifurcation point $K_2\simeq0.413765$. According to this analysis, the bifurcation 
appears to be a supercritical Hopf bifurcation, which gives rise to a new attractor:
the T2 dynamics indicated in Fig.~\ref{fig:phasediagram}.  Nevertheless, as it can be read from the  phase diagram,
in this new regime the limit cycle does not seem to correspond to a rotation around the unstable fixed point.
A detailed analysis of the dynamics of the system close to the bifurcation point shows that the
amplitude of the oscillations grows as $\sqrt{K-K_2}$ whereas the frequency does not change significantly.
We conjecture that the effect is due to the presence of infinitely many nearly marginal directions, 
which induce a detachment of the T2 attractor from the plane spanned by the unstable directions 
of the fixed point.

\section{Increasing dynamical complexity}\label{sec:chaos}

An analytical study of time-dependent nonlinear regimes is typically unfeasible. So, from now on,
we proceed exclusively on a numerical basis, by relying on the integration of both microscopic and 
macroscopic equations.
The Poincar\'e section is a qualitative but informative tool to understand the dynamical properties of
the different regimes observed for larger values of $K$.
It was already used in~\cite{nakagawa1995}. Here we consider different observables for reasons that will be
clear in a moment. More precisely, we introduce the collective variables
\begin{equation*}
Z_n^{(k)}=\left|\int_0^{2\pi}d\psi R(\psi,t_n)P(\psi,t_n)\mathrm{e}^{ik\psi}\right|
\end{equation*}
where $t_n$ is the time at which $|\overline z|$ reaches a local maximum. This definition is basically
an extension of the order parameters typically used to characterize phase oscillators, where the
radial contribution $R$ is, for obvious reasons, absent.
In phase oscillators, the $Z^{(k)}$ parameters are functionally related to one another
whenever the Ott-Antonsen Ansatz is valid~\cite{Ott-Antonsen-2008}. 
It is therefore instructive to look at their mutual relationship.

\begin{figure}
	\includegraphics[width=0.5\textwidth]{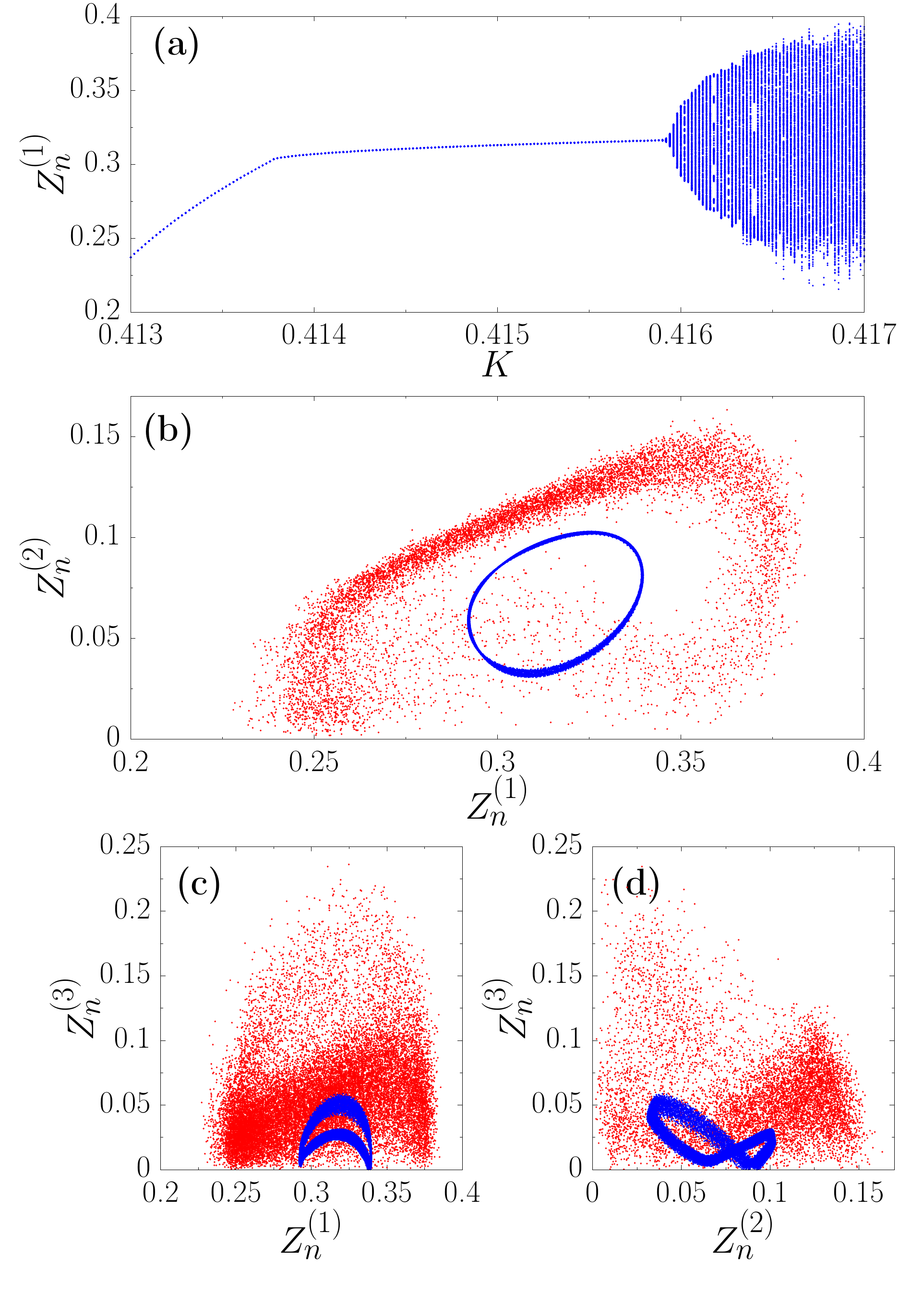}
	\caption{
		(a) Poincar\'e map of $Z_n^{(1)}$ for different values of the coupling strength $K$ computed using macroscopic simulations.
		The T1 and T2 regimes appear as stable fixed points, and T3 emerges through a Hopf-bifurcation.
		Simulations computed along $2\times 10^4$ time units after discarding a transient of $5\times10^5$ time units.
		(b-d) Poincar\'e sections resulting from macroscopic simulations for $K=0.416$ (blue dots) and $K=0.4165$ (red dots).
		Simulations computed along $10^5$ time units after discarding a transient of $5\times10^4$ time units.
		 }
	\label{fig:poincare}
\end{figure}

In Fig.~\ref{fig:poincare}(a), we plot $Z_n^{(1)}$ versus $K$.
Since the Poincar\'e section reduces by one unit the dimensionality of the underlying attractor,
a periodic collective dynamics manifests itself as a single point for a given $K$ value.
This is indeed what we see for $K<K_3$, although we should notice that the initial section of
the curve corresponds to SCPS, where the order parameter is strictly constant. The fuzzy region covered for
$K>K_3$ corresponds to a tiny interval where a T3 dynamics initially unfolds, followed by a
chaotic regime, analyzed in more quantitative way in the second part of this section.

The lower panels of Fig.~\ref{fig:poincare} contain various Poincar\'e section in the T3 
region (blue dots) and in the chaotic region (red dots). 
The points have been obtained by integrating the macroscopic equations. Analogous pictures have been
obtained by integrating 32768 oscillators, but significantly blurred by finite-size 
effects\footnote{The obfuscation of blue points clearly visible in panels (c) and (d) of
Fig.~\ref{fig:poincare} are also artifacts due to the finite accuracy in the determination
of the Poincar\'e section}.
The main message that we learn by comparing the three Poincar\'e sections is that there is no functional dependence
among the first three order parameters, thus suggesting that they are really independent variables --
an indirect evidence that the Watanabe-Strogatz theorem does not apply to this context.

How irregular is the regime which settles beyond $K_4$? How does chaos emerge?
Obtaining a clean answer turned out to be arduous.
In fact, one has to deal with:
(i) long transients; (ii) the spontaneous formation of metastable states;
(iii) finite-size effects (in microscopic simulations); (iv) the sporadic formation of 
highly localized, cluster-like structures (in macroscopic simulations). All of them required much care
and long lasting simulations.

\begin{figure}
	\includegraphics[width=0.5\textwidth]{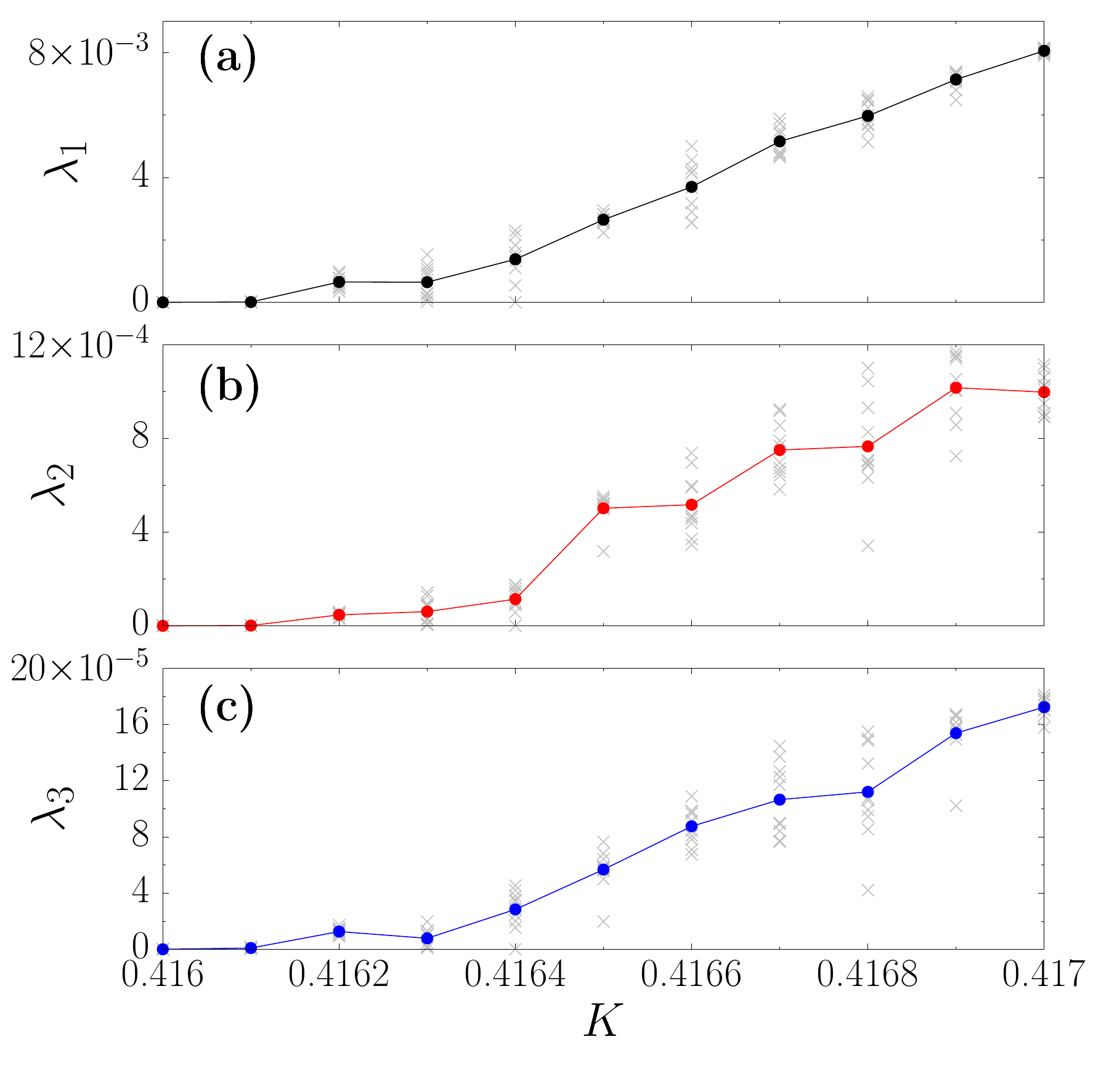}
	\caption{
		The three largest Lyapunov exponents of the microscopic system computed over different values of the coupling strength $K$.
		Circles indicate the average over ten different realizations, each starting from a different initial condition. Gray crosses show the outcome of the different single realizations. Each simulation consists of $N=4096$ oscillators, with a quenched disorder of the order of $10^{-14}$. The total computation time is of $10^7$ time units, of which $10^{5}$ are transient. Simulations computed using fourth order Runge-Kutta with time step $dt=0.01$. The Gram-Schmidt orthogonalization is invoked every 10 time steps.
		}
	\label{fig:micro_le}
\end{figure}

We first focus on the computation of the standard, microscopic, Lyapunov exponents.
$N=4096$ turns out to be sufficiently large to ensure negligible finite-size corrections.
The convergence is nevertheless very slow and a good way to cope with it is
by launching simulations from different initial conditions.
Moreover, we also add a small heterogeneity of the order of $10^{-14}$ among the oscillators
in order to prevent the formation of spurious clusters due to the finite floating point representation.
The parameter dependence of the first three Lyapunov exponents is summarized in Fig.~\ref{fig:micro_le}.
The three exponents appear to become positive for approximately the same coupling strength 
$K_4\approx 0.4161\ldots$. This is a first indication that we are before a new transition to chaos. 
It differs from the typical transitions to low-dimensional chaos (period doubling, intermittency, quasiperiodicity),
which are accompanied by the change of sign of a single exponent.
On the other hand, there is no similarity with the transitions expected in high-dimensional systems, 
such as spatio-temporal intermittency, where a bunch of exponents becomes positive, whose numerosity
is proportional to the system size~\cite{Chate-Manneville-87}: a typical signature of extensivity.

\begin{figure}
	\includegraphics[width=0.5\textwidth]{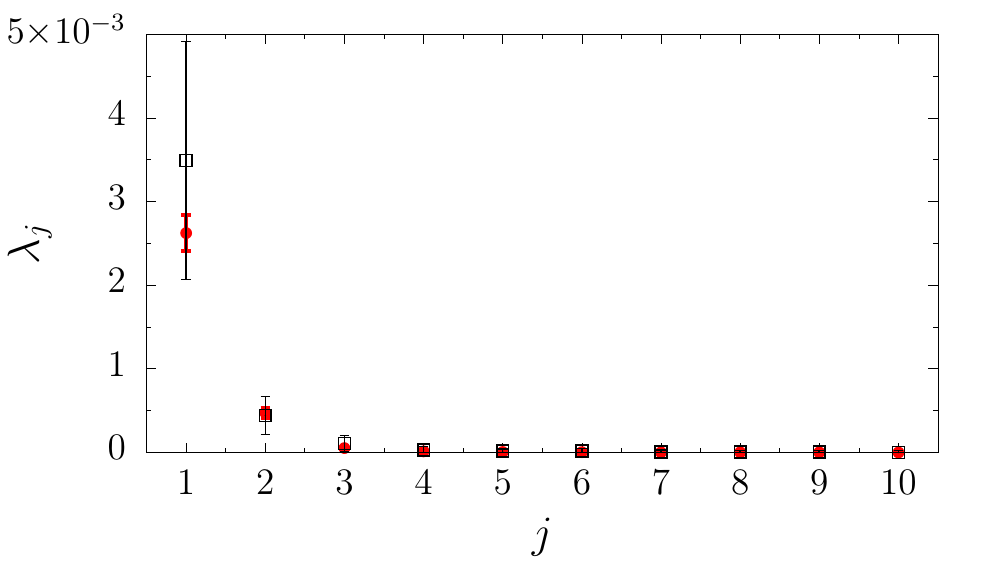}
	\caption{
		The 10 largest Lyapunov exponents for $K=0.4165$ computed using microscopic (red circles)
		and macroscopic (open black squares) simulations.
		The different points show the average over 10 (microscopic) and 50 (macroscopic) simulations starting from different initial conditions, with error bars indicating the standard deviation.
		Macroscopic simulations computed  over $2\times 10^4$ time units after discarding a transient of $5\times 10^3$.
		Parameters for microscopic simulations as in figure~\ref{fig:micro_le}.
	}
	\label{fig:lyap}
\end{figure}

A more detailed representation of the overall degree of instability is given in Fig.~\ref{fig:lyap},
where we plot the first 10 Lyapunov exponents deeply inside the chaotic region. 
All exponents beyond the third one are practically equal to zero.
As a result, by virtue of the Kaplan-Yorke formula, the underlying attractor
is characterized by a large (possibly infinite in the thermodynamic limit) dimension,
in spite of the presence of just three positive exponents: an additional reason to classify this 
regime as genuinely new.

How does the microscopic instability compare to the macroscopic dynamics? 
In Fig.~\ref{fig:lyap}, we plot also the macroscopic Lyapunov exponents, obtained
by linearizing the evolution equations (\ref{eq:attractor2},\ref{eq:probability},\ref{eq:coupling1})
along a generic trajectory.
In spite of the large fluctuations (especially those affecting the maximum exponent), the macroscopic spectrum 
is very similar to the microscopic one.
This correspondence is far from obvious and shall be addressed in the final part of this section.
Here, we want to stress that the presence of positive macroscopic exponents shows that we
are before a form of collective chaos.

In all mean-field models so far investigated in the literature, collective chaos is accompanied by 
the instability of the single dynamical units, which can be quantified by interpreting the
mean field as an external driving force and thereby determining the so-called transverse Lyapunov 
exponent $\lambda_T$.
This is the case of logistic maps~\cite{ShibataKaneko1998} as well as of
Stuart-Landau oscillators in a different parameter region~\cite{Takeuchi-Chate2013}. From now on,
we refer to this regime as to {\it standard collective chaos} (SCC).
We now show that the collective dynamics observed beyond $T3$ has a different nature.

In the present context, $\lambda_T$ can be determined by linearizing the evolution
equation (\ref{eq:coupled-oscillators}) under the assumption that $\bar z(t)$ is an external forcing,
\begin{equation}\label{eq:transverse}
	\dot u=[1-K(1+ic_1)]u-(1+ic_2)z(2z^* u+zu^*) \; ,
\end{equation}
where $u$ denotes an infinitesimal perturbation of $z$. This equation being two-dimensional
($u$ is a complex variable) is characterized by two (transverse) Lyapunov exponents.
The lower exponent is unavoidably negative (in the present case it expresses the stability of deviations 
from the time-dependent curve $\mathcal{C}$).
Less trivial is the value of the largest transverse exponent $\lambda_T$, which quantifies the stability
of perturbations aligned with the tangent to the curve ($\mathcal{C}$).
The best way to describe the underlying phenomenology is by invoking the standard multifractal formalism,
which takes into account the fluctuations of the Lyapunov exponents (see, e.g., the book~\cite{Pikovsky-Politi-2016}).
Let us start by introducing the generalized Lyapunov exponent
\begin{equation*}
\mathcal{L}(q) = \lim_{\tau\to\infty} \frac{1}{q\tau} \ln \langle |H(\tau)u|^q \rangle
\end{equation*}
where $H$ represents the Jacobian integrated over a time $\tau$ (from Eq.~(\ref{eq:transverse})).
$\mathcal{L}(0)$ corresponds to the standard the Lyapunov exponent, while $\mathcal{L}(1)$ corresponds
to the topological entropy  (in case there is a single positive exponent)~\footnote{We warn the reader that
a different definition is often found in the literature, where $q=1$ corresponds to the standard Lyapunov
exponent. Here we follow the same notations as in \cite{Pikovsky-Politi-2016}}.
In fact, $\mathcal{L}(1)$ yields the expansion rate of an {\it arc} of initial conditions initially
aligned along the most expanding direction.
In the current context, the orientation of the arc corresponds to that of the curve $\mathcal{C}$.
Since the curve itself has a fluctuating but finite length, the length of any subsegment does neither
grow nor diverge in time, so that $\mathcal{L}(1)=0$.

The generalized Lyapunov exponents can be determined from the
probability $\mathcal{P}(\Lambda,\tau)$ to observe a finite-time Lyapunov exponent $\Lambda$ over
a time $\tau$ and thereby introducing the large deviation function $S(\Lambda)$ 
\begin{equation*}
S(\Lambda) = \lim_{\tau\to\infty} - \frac{\ln \mathcal{P}(\Lambda,\tau)}{\tau} \; .
\end{equation*}
The function $S(\Lambda)$ is equivalent to $\mathcal{L}(q)$, the connection between the two
representations being given by the Legendre-Fenchel transform~\cite{Pikovsky-Politi-2016}
\[
q\mathcal{L}(q) = q \Lambda^* - S(\Lambda^*)  \; ,
\]
where $q= S'(\Lambda^*)$.
In the Gaussian approximation
\[
S(\Lambda) = \frac{(\Lambda-\lambda_T)^2}{2D}
\]
where $\lambda_T$ is the standard transverse Lyapunov exponent, while $D$ is the corresponding
diffusion coefficient defined as
\[
D = \lim_{\tau\to\infty} \tau \left (\overline{\Lambda(\tau)^2}-\lambda_T^2 \right) \; .
\]
As a result, we eventually find that
\begin{equation}\label{eq:finaltran}
\mathcal{L}(1) = \lambda_T + \frac{D}{2} \; .
\end{equation}
Both $\lambda_T$ and $D$ can be determined from the time evolution of 
$\gamma(\tau)=\tau \Lambda(\tau)$\footnote{Here, $\Lambda(\tau)$ denotes the 
finite-time Lyapunov exponent computed by following a single trajectory from time
0 to time $\tau$.}.
In fact, $\gamma(\tau)$ is the logarithm of the expansion factor over a time $\tau$; it is basically a
Brownian motion with a drift velocity $\lambda_T$ and a diffusion coefficient $D$.
For $K=0.4165$, upon integrating over $10^7$ time units we find a slightly negative Lyapunov exponent 
$\lambda_T\approx (-1.7\pm 0.6) \times 10^{-5}$, while $D \approx (4.4\pm 0.3) \times 10^{-5}$.
As a result, from Eq.~(\ref{eq:finaltran}), 
$\mathcal{L}(1) \approx (0.5 \pm 0.75) \times 10^{-5}$, a value compatible with the expected vanishing exponent.
In other words, we see that the fluctuations of the finite-time Lyapunov exponent 
compensate the slightly negative $\lambda_T$ and ensure a vanishing expansion factor for
the curve length.

Altogether, the message arising from the multifractal analysis is that the (unavoidable) 
fluctuations of the transverse Lyapunov exponent induce a set of singularities
for the corresponding probability density (in the regular SCPS, there are no fluctuations of
the Lyapunov exponent, which is identically equal to 0).

A posteriori, this observation accounts for the difficulties encountered in our simulations of the chaotic 
phase: in fact, the formation of temporary clusters which so much affect the accuracy of our simulations, 
irrespective whether they are carried out at the microscopic or macroscopic level, are nothing but
a manifestation of the unavoidable presence of singularities, which are intrinsically associated to
the self-sustainment of a fluctuating probability density.

We conclude this section by commenting on the similarity between macroscopic and microscopic Lyapunov spectra.
The correspondence is unexpected since they arise from two different descriptions of the world. 
In the microscopic approach, the key variables are the ``positions'' of the single
oscillators, while the macroscopic approach deals with their distribution in phase space.
Imagine, for simplicity, to deal with $N$ particles constrained to move along a given curve of fixed length:
a virtually infinitesimal microscopic perturbation corresponds to a shift of each particle over a scale
that is by definition small compared to the interparticle distance, of the order of $1/N$. 
On the other hand, to meaningfully interpret a perturbation of the positions
as a perturbation of the corresponding probability density, it must occur on a scale larger 
than the statistical fluctuations, which is of order $\sqrt{1/N}$. A priori, there is no
guarantee that the linearization of the microscopic equations still hold over such ``large scales" 
(see Ref.~\cite{PolitiPikovskyUllner2017}, for a more detailed discussion of this point).
In fact, in typical instances of SCC, macroscopic and microscopic Lyapunov spectra substantially
differ from one another.
In the toy model of SCC discussed in~\cite{PolitiPikovskyUllner2017}, the collective
dynamics is characterized by a single positive macroscopic exponent, while the number of positive exponents 
is proportional to the number of oscillators, in the microscopic dynamics.
The main point of discussion is whether and when some of the microscopic exponents ``percolate" to the 
macroscopic level.  In Ref.~\cite{Takeuchi-Chate2013} it is conjectured that this happens whenever 
the corresponding covariant Lyapunov vector has an extensive nature, but it is still unclear 
under which conditions this opportunity materializes. 

So, what is the difference with the collective chaos discussed in this paper?
The stability analysis of splay states in leaky integrate-and-fire (LIF) neurons
can help to shed some light. At the collective level, the state splay corresponds to
a trivial staionary homegeneous distribution and its stability can be studied by diagonalizing the
corresponding linearized evolution operator. This step was already performed 
in the early'90s, determining an analytical expression of the entire spectrum in the weak 
coupling limit~\cite{Abbott1993}. 
More recently, the same problem was revisited from the microscopic point of view, analyzing
an arbitrary number $N$ of neurons~\cite{OlmiPolitiTorcini2014},
finding that the leading exponents progressively approach the macroscopic ones (upon increasing $N$),
analogously to what observed in Fig.~\ref{fig:lyap}.
The main difference between splay states and SCC is the absence of microscopic chaos and thereby
the absence of the statistical fluctuations of size $1/\sqrt{N}$ which would otherwise represent a sort of 
``barrier'' separating the microscopic from the macroscopic world 
(see~\cite{PolitiPikovskyUllner2017} for additional considerations of this point).
The maintanance of ordering observed in the collective chaos discussed in this paper, makes it
closer to the splay state than to SCC.


\section{Conclusions and open problems}\label{sec:conclusions}

In this paper we have developed a formalism that allows characterizing an intermediate regime
where amplitude oscillators keep some phase-oscillator properties (i.e. the
alignment along a smooth curve), while starting to exhibit nontrivial amplitude oscillations.
Our analysis deals with Stuart-Landau oscillators, but nothing precludes the application of the same
formalism to generic oscillators, in so far as their evolution remains confined 
to a smooth curve $\mathcal{C}$.

We show that SCPS is a generic regime: a necessary condition for it to be self-sustained
is the presence of more than one Fourier harmonics in the effective coupling function, a property that
is here induced by the amplitude dynamics.
Our formalism allows for an almost analytical characterization of SCPS and, in particular,
to determine the bifurcation point, beyond which complex time-dependent states arise.
If SCPS is by itself a non-intuitive regime, since the single oscillators behave quasi-periodically 
without displaying any locking phenomena, the chaotic SCPS described in section~\ref{sec:chaos}
is even more so. Each oscillator, under the action of the self-sustained chaotic mean-field, 
is consistently marginally stable 
(the transversal generalized Lyapunov exponent being equal to zero for $q=1$)
when the coupling strength is varied.
An implication of this observation is that the probability density is unavoidably characterized by the
presence of singularities that manifest themselves as temporary clusters.

If and when the transversal Lyapunov exponent becomes positive, a transition to SCC occurs,
accompanied by the divergence of the curve $\mathcal{C}$, which would thereby ``fill'' the phase space
(in a fractal way).
In the parameter range explored in this paper, this transition is preceded by the onset of a 
non-conventional incoherent state (\emph{i.e.} nonuniform distribution characterized by a zero order parameter), 
which restores a perfectly circular shape of $\mathcal{C}$. 
It will be worth clarifying the possibly universal mechanisms that may lie behind such a kind of transition.

Finally, the onset of a chaotic SCPS is itself an entirely new phenomenon which involves the 
simultaneous emergence of more than one positive Lyapunov exponent (actually it looks like three of them). 
While we could imagine simple mechanisms for the emergence of discontinuous changes (see, e.g.
attractor crises), the justification of a continuous transition such as the one discussed in
this paper is by far more intriguing. Last but not least, the question whether chaotic SCPS
can be observed in perfect phase oscillators remains open. 

\section*{Acknowledgement}
This work has been financially supported by the EU project COSMOS (642563).
We wish to acknowledge Ernest Montbri\'o for early suggestions and 
enlightening discussions.

\appendix

\section{First order approximation of SCPS}\label{sec:app2}

In this section we develop a perturbative approach to determine the stationary solutions of 
Eqs.~(\ref{eq:scpsR},\ref{eq:scpsP}) close to the transition to SCPS, $K\sim K_1$.
The main idea is, as usual, the identification of the leading terms.
Slightly above $K_1$, the shape $R_0$ of the attractor and the corresponding
density distribution $P_0$ of the phases are close to the splay state, so that we can write
\begin{eqnarray}
R_0(\theta)&=&\sqrt{1-K}+r(\theta) \label{eq:firstorder} \\
P_0(\theta)&=&\frac{1}{2\pi}+p(\theta) \; .\nonumber
\end{eqnarray}
Expanding the self-consistent condition~(\ref{eq:self-consistent}) up to linear terms, we obtain 
\begin{align}\label{eq:perturbationz}
\bar z&=\int_0^{2\pi}d\psi P(\psi)R(\psi)e^{i\theta}\nonumber \\
&=\frac{1}{2\pi} \int _0^{2\pi}e^{i\psi}d\psi r(\psi)+\sqrt{1-K}\int_0^{2\pi}d\psi p(\psi)\mathrm{e}^{i\psi}\nonumber\\
&=\frac{1}{2\pi}\tilde r(1)+\sqrt{1-K}\tilde p(1)\;.
\end{align}
By then expanding equation~(\ref{eq:scpsR}) up to first order, one obtains
\begin{equation*}
[r(\theta)]_\theta=\frac{2(1-K)r-K\bar z \sqrt{1+c_1^2} \cos(\nu-\theta)}{\Delta} \; ,
\end{equation*}
where
\begin{align*}
\Delta:=&\omega+c_1K+c_2(1-K)\;, \\
\nu:=&\arctan c_1
\end{align*}
By introducing the notations
\begin{align*}
a_\xi\mathrm{e}^{i\xi}:=& \frac{2(1-K)}{\Delta} - i, \\
A:=& \frac{K \sqrt{1+c_1^2}}{a_\xi\Delta},\\
\end{align*}
one can write the solution of the ODE as
\begin{equation}\label{eq:perturbationr}
r(\theta)=\overline zA\cos(\nu+\xi-\theta) \; .
\end{equation}

Accordingly, 
\begin{equation*}
\tilde r(1) = \pi \overline z A \mathrm{e}^{i(\nu+\xi)}
\end{equation*}
By expanding Eq.~(\ref{eq:scpsP}) in the same way, at zero order we obtain
\begin{equation*}
\eta = -\frac{\Delta}{2\pi}
\end{equation*}
while, at first order, 
\begin{align*}
p(\theta)=\frac{-2c_2(1-K) r(\theta)+K\bar z\sqrt{1+c_1^2}\sin(\nu-\theta)}{2\pi\Delta\sqrt{1-K}}
\end{align*}
By replacing the expression for $r(\theta)$,
\begin{align*}
p(\theta)=C \bar z \left [ 
-\frac{2c_2(1-K)}{a_\xi\Delta} \cos(\nu+\xi-\theta) + \sin(\nu-\theta) \right ]
\end{align*}
where
\begin{equation*}
C =  \frac{K}{2\pi\Delta} \sqrt{\frac{1+c_1^2}{1-K}}\;.
\end{equation*}
By finally, recombining the two sinusoidal terms, we find
\begin{align*}
p(\theta)=& a_\gamma C \bar z \cos(\nu+\gamma-\theta)
\end{align*}
where 
\begin{equation*}
a_\gamma \mathrm{e}^{i\gamma} =  \frac{-2(1-K)c_2 \mathrm{e}^{i\xi}}{a_\xi\Delta} - i\;.
\end{equation*}
Thus, also $p(\theta)$ is a purely harmonic function and
\begin{equation*}
\tilde p(1) = \pi \overline z a_\gamma C \mathrm{e}^{i(\nu+\gamma)} \; .
\end{equation*}
The overall effect of the coupling is finally determined by inserting the expressions for $\tilde r(1)$ and
$\tilde p(1)$ into Eq.~(\ref{eq:perturbationz}). Since both terms are proportional to $\bar z$, we can
interpret
\begin{equation*}
G := g_z\mathrm{e}^{i\delta} := \frac{A}{2} \mathrm{e}^{i(\xi+\nu)} + \pi a_\gamma C\sqrt{1-K} \mathrm{e}^{i(\gamma+\nu)} \; ,
\end{equation*}
as the expansion factor of $\bar z$ in the presence of a given small modulation of both $\mathcal{C}$ and the probability density.
Imposing $G=1$ finally allows determining a self-consistent solution. More precisely, one can determine
the frequency $\omega$ of the collective rotation (so far unspecified) and the bifurcation point $K$.
For $c_1=-2$ and $c_2=3$, we obtain $K=K_1\simeq 0.4123...$ and $\omega=\omega_c=-2.5261...$, \emph{i.e.}, in agreement
with the stability analysis of the splay state.\\

\section{Stability of SCPS} \label{sec:app1}

The coefficients of the linearized equations (\ref{eq:scpsdotr},\ref{eq:scpsdotp}) are
\begin{align*}
F^{(v)}(\theta):=&(1-K)-3R_0(\theta)^2+2c_2 R_0'(\theta)R_0(\theta)\\\qquad&\quad
+\frac{K \overline z R_0'(\theta)}{R^2_0(\theta)}\sqrt{1+c_1^2}\sin(\nu-\theta)\;,\\
G^{(v)}(\theta):=& Kc_1+\omega+c_2 R_0^2-\frac{K\overline z}{R_0(\theta)} \sqrt{1+c_1^2}\sin(\nu-\theta)\;,\\
X^{(v)}(\theta):=&\frac{K\sqrt{1+c_1^2}}{2}\left(1-\frac{R_0'(\theta)}{iR_0(\theta)}\right)e^{i(\nu-\theta)}\;,\\
Y^{(v)}(\theta):=&\frac{K\sqrt{1+c_1^2}}{2}\left(1+\frac{R_0'(\theta)}{iR_0(\theta)}\right)e^{-i(\nu-\theta)}\;,\\
F^{(u)}(\theta):=&2c_2 P_0(\theta)R_0(\theta)+\frac{K  P_0(\theta)\overline z}{R_0(\theta)^2}\sqrt{1+c_1^2}\sin(\nu-\theta)\;,\\
G^{(u)}(\theta):=& Kc_1+\omega+c_2 R_0(\theta)^2-\frac{K\overline z}{R_0(\theta)} \sqrt{1+c_1^2}\sin(\nu-\theta)\;,\\
X^{(u)}(\theta):=&-\frac{KP_0(\theta)}{2iR_0(\theta)}\sqrt{1+c_1^2}e^{i(\nu-\theta)}\;,\\
Y^{(u)}(\theta):=&\frac{KP_0(\theta)}{2iR_0(\theta)}\sqrt{1+c_1^2}e^{-i(\nu-\theta)}\;.
\end{align*}

The equations are better solved in Fourier space. By invoking the Fourier transform, 
the integrals in the linearized mean-field expression~(\ref{eq:selfconsistent})
can be expanded as a linear combination of $\{\tilde v(k,t)\}_{k=-\infty}^{\infty}$ 
and $\{\tilde u(k,t)\}_{k=-\infty}^{\infty}$,
\begin{equation*}
w(t)=\frac{1}{2\pi}\sum_{k=-\infty}^\infty \tilde v(k)\tilde P_0(1-k)+\tilde u(k)\tilde R_0(1-k)
\end{equation*}
and
\begin{equation*}
\hat w(t)=\frac{1}{2\pi}\sum_{k=-\infty}^\infty \tilde v(k)\tilde P_0(-1-k)+\tilde u(k)\tilde R_0(-1-k).
\end{equation*}
Similarly, we express Eqs.~(\ref{eq:scpsdotr}) and~(\ref{eq:scpsdotp})
in Fourier space.
Retaining the terms for each wavelength $k$ we obtain 
an infinite system of linear equations,
\begin{align*}
[ \tilde v(k)]_t&= \frac{1}{2\pi}\sum_{j=-\infty}^{\infty} \tilde v(j)
\Bigl[
\tilde F^{(v)}(k-j)-ij\tilde G^{(v)}(k-j)\\&+
\tilde X^{(v)}(k)\tilde P_0(1-j)+\tilde Y^{(v)}(k)\tilde P_0(-1-j)
\Bigr]\\
&+
\tilde u(j)
\Bigl[
\tilde X^{(v)}(k)\tilde R_0(1-j)+\tilde Y^{(v)}(k)\tilde R_0(-1-j)
\Bigr]
\end{align*}
and
\begin{align*}
[ \tilde u(k)]_t&= \frac{-ik}{2\pi}\sum_{j=-\infty}^{\infty} \tilde v(j)
\Bigl[
\tilde F^{(u)}(k-j)\\&+
\tilde X^{(u)}(k)\tilde P_0(1-j)+\tilde Y^{(u)}(k)\tilde P_0(-1-j)
\Bigr]\\
&+
\tilde u(j)
\Bigl[
\tilde G^{(u)}(k-j)+
\tilde X^{(u)}(k)\tilde R_0(1-j)\\&+\tilde Y^{(u)}(k)\tilde R_0(-1-j)
\Bigr]\;.
\end{align*}
\\

\nocite{*}
\bibliography{references}

\end{document}